\documentclass{article}
\usepackage{amsmath,epsfig,amssymb}

\begin{document}

\title{Breakdown of Bell's Theorem for incompatible measurements}

\author{Karl Hess }

\date{ Beckman Institute for Advanced Science and Technology, Departments
of Electrical and Computer Engineering and Physics and
Center for Advanced Study, University of Illinois, 405 N. Mathews
Avenue, Urbana, Illinois 61801, USA
 }


\maketitle

\begin{abstract}

Bell's theorem contains the proposition that the
Einstein-Podolsky-Rosen (EPR) theory (hypothesis) of the existence
of elements of reality together with Einstein locality permits a
mathematical description of EPR experiments by functions that are
all defined on one common probability space. This proposition leads
in turn to restrictions for possible experimental outcomes that Bell
expressed in terms of his well known inequalities and that Vorob'ev
and others had investigated before Bell. Summarizing several
previous publications and adding new material, the above proposition
is refuted by Einstein-local counterexamples from classical physics
and shown to involve additional assumptions that can not be
justified for mutually exclusive (incompatible) measurements and
experiments. Moreover, criticism of our work by Mermin who invoked
``standard sampling arguments" is shown to be false.

\end{abstract}

\section{Introduction}

It was shown in \cite{hppnas} that Bell's ``no-go proof", a complex
of physical and mathematical statements that are now known as Bell's
theorem (published in \cite{bell} and extended in \cite{bellbook}),
contains the unjustified and incorrect assumption that all functions
describing EPR experiments, even incompatible experiments, must be
definable on one common probability space, and that this assumption
does not even permit to cover all of the space of variables and
functions of classical physics. It was the authors conviction that
the criticism of \cite{hppnas} by Mermin \cite{mermin},
\cite{mermin1} had been completely refuted, in \cite{hpfound},
\cite{hpwqi}, \cite{hpgerman} and \cite{hpleiden}. In addition, a
direct logical contradiction of Bell's assumption to the framework
of classical physics (special relativity) was demonstrated and
published in \cite{hpfound} and \cite{hpleiden}. Only recently was
the author informed by A. J. Leggett that it was Leggett's believe,
and that of 90 \% of the experts, ``that the work of the late John
Bell on what is sometimes known as ``quantum non-locality" is one of
the most profound results of physics" \cite{legpriv0} and that
Mermin's arguments in \cite{mermin}, \cite{mermin1} did, in
Leggett's opinion, carry the day and that ``he (Mermin) puts the
point at least as clearly as I (Leggett) could" \cite{legpriv}. The
author therefore has summarized and clarified the reasoning that was
presented in \cite{hpfound}, \cite{hpwqi}, \cite{hpgerman} and
\cite{hpleiden} in section 3 and related this reasoning directly to
Mermin's criticism in section 4. New explanatory material in terms
of vector valued stochastic processes is also added. The author is
convinced that this present manuscript eclipses those of Mermin
\cite{mermin}, \cite{mermin1} in clarity and shows that Mermin's
reasoning is based on a lack of mathematical rigor and generality
and can not be brought into correspondence with Kolmogorov's
probability theory as well as its precise relationship to actual
experiments. All criticism of the theorem of Bell by the author and
W. Philipp therefore stands and new criticism and counterexamples
are added in sections 3 and 4.

The author likes to emphasize that the schism among physicists about
quantum non-locality goes significantly beyond the opinion of a
negligible minority. This fact is demonstrated by the recent work of
Gerard 't Hooft \cite {hooft}. Kerson Huang \cite{huang} writes in
his elementary text: ``...local gauge invariance frees us from the
last vestige of action at a distance...". It is impossible to
include references to all relevant work, and we list in addition to
the other references in this paper just \cite{khrbook} and
\cite{morgan2} as further examples.

It also should be noted that the author subscribes fully to the
teachings of both quantum and Kolmogorov probability (as different
and well proven probability frameworks) and to their well known
relationship to actual experiments (see e.g. \cite{brpe}). The
author has neither any criticism for these frameworks nor for the
definition of the ``elements of physical reality" of the EPR paper
\cite{EPR} nor for the EPR-type experiments performed by Aspect
\cite{Aspect} and others. The author criticizes exclusively the work
of Bell as not being general enough to apply to general physics
problems (quantum and/or classical) and the work of Bell's followers
for the same reason and for actual logical and mathematical
mistakes.

\section{Consistent random variables and probability spaces}

Probability theory as applied to problems of science is and must be
more than a mathematical game \cite{willi}. It is therefore
necessary to relate the axiomatic system of probability theory, and
we concentrate on that of Kolmogorov, in a unique logical way to the
actual experiments. Probability theory becomes then and only then a
scientific tool and is sometimes called a ``pre-statistics". The
unique and logical connection of Kolmogorov probability to actual
experiments is commonly given by the introduction of a sample space
denoted by upper case Greek symbols such as $\Omega$ \cite{willi}.
The sample space contains elements denoted by lower case Greek
symbols such as $\omega$ that have a unique and logical relation to
experiments. As an example consider a roulette type of game that can
result in even ($e$) or odd ($o$) outcomes. Taking the ``composite"
experiment of obtaining two results in a row we have (see also
\cite{willi}):
\begin{equation}
\Omega = \{ee, eo, oe, oo \} \label{j08na1}
\end{equation}
were we have used an explicit notation for the even-odd sequences
that are possible. The goddess of fortune, Tyche makes then one
choice for a particular experiment, in this case a composite or
``multiple stage" experiment consisting of two trials, by choosing a
point $\omega^{act}$ of $\Omega$ and thus ``crystallizing into
existence" one sample point e.g. $eo$ which, of course, is revealed
in two stages. It is obvious at this point that in order to create a
general mathematical model we must include all the sample points of
Eq.(\ref{j08na1}) and can not just leave out $oo$ or $oe$ without
loosing the capability of our pre-statistics to make statements
about the actual experiments. When deciding on a sample space we
need to remember Fellers \cite{feller} word:``If we want to speak
about experiments or observations in a theoretical way and without
ambiguity, we first must agree on the simple events representing the
thinkable outcomes; {\it they define the idealized experiment...}.
By definition every indecomposable result of the (idealized)
experiment is represented by one, and only one, sample point. The
aggregate of all sample points will be called the sample space". We
consider in the following, for reasons of physical clarity, only
countable sets $\Omega$. The collection of all subsets of $\Omega$
forms then the set $F$ of events of probability theory. In order to
define a probability space we still need to introduce a measure $P$
on $F$ which assigns to each event (subset of $\Omega$) a real
number of the interval $[0, 1]$ with $P(\Omega) = 1$. This
assignment needs to be done also in a fashion that is logical with
respect to the science-problem to which we apply probability theory.
We then have formed a probability space $(\Omega, P)$.

For applications of Kolmogorov's probability theory it is desirable
and advantageous to introduce random variables to describe certain
possible experimental outcomes. {\it Random variables are functions
on a probability space} and we denote them with upper case Roman
letters such as $X$, $Y$ as opposed to the actual experimental
outcomes for which we use lower case Roman $x, y$. The range of the
functions (random variables) is chosen to correspond to the range of
experimental outcomes, e.g. to the discrete (because of assumed
countability) eigenvalue spectrum of certain quantum operators that
in turn correspond to some measurement equipment. For example we may
denote the outcomes of spin measurements by $A_{\bf a} = \pm 1$ or
$B_{\bf b} = \pm 1$ and characterize the measurement equipment (e.g.
magnets) by bold faced subscripts such as $\bf a, \bf b$. Vector
random variables such as described by a pair $(A_{\bf a}, B_{\bf
b})$ may also be used. The domain of the random variables is the
probability space. The random variables are only fully determined
when both range and domain are given and clearly linked to the
actual experiments and the science of a given application. This
opens the problem that is discussed extensively in this paper. Any
given application will have some non-trivial restrictions on the
choice of functions, their range and their domain, and certain
functions may just not be defined on the same domain. We start to
illustrate this by examples.

Consider the above roulette game and random variables $X(\omega),
Y(\omega)$ that can assume the values $e, o$. The pair outcomes of
Eq.(\ref{j08na1}) can then also be described by the pair of random
variables $(X(\omega),Y(\omega))$ and the use of and consistency of
the sample space is clear as one can see by inserting a possible
actual choice of Tyche such as $\omega^{act} \equiv eo$ with
$X(\omega^{act}) = e$ and $Y(\omega^{act}) = o$ etc..

As a more complicated example consider a game that is not ``fair"
but is influenced by some hidden magnetic substance that reacts to a
magnetic machinery of which we only know some ``orientation" given
by unit vectors such as $\bf a$ or $\bf b$, influencing the first
experiment of each pair and another hidden magnetic machinery
characterized by unit vectors $\bf b$ and $\bf c$ influencing the
second experiment of the pair. We now describe the outcome of the
first experiment by the random variables $X_{\bf a}(\omega)$,
$X_{\bf b}(\omega)$  and the second by $Y_{\bf b}(\omega)$, $Y_{\bf
c}(\omega)$. We still can, as we will see in the lemma of section 3,
describe pair outcomes by pairs of random variables such as $(X_{\bf
a}(\omega),Y_{\bf b}(\omega))$. However, if we try to create more
complicated composites such as $(X_{\bf a}(\omega),X_{\bf b}(
\omega),Y_{\bf b}( \omega),Y_{\bf c}( \omega))$ we need to ask
ourselves if the settings are not mutually exclusive. For example,
there may only be one single magnet available for the first two
experiments as well as another magnet for the second two. Then it is
not clear that Tyche will be able to find an $\omega^{act}$ that
``crystallizes" $(X_{\bf a}(\omega^{act}),X_{\bf b}(
\omega^{act}),Y_{\bf b}( \omega^{act}),Y_{\bf c}( \omega^{act}))$
into existence \cite{willi} without physical or mathematical
contradictions. A particular danger to overlook this fact is given
when the experimental outcomes are characterized just by numbers
e.g. $X, Y = \pm 1$ and one is therefore tempted to perform
algebraic operations such as additions, subtractions and
multiplications. One may also apply these operations to random
variables. However, if no $\omega^{act}$ exists that assigns values
to all of them, this can not necessarily be done. As we will see, it
is not always certain that one can find one common probability space
on which a number of functions are defined and thus become random
variables. In order to describe different (particularly mutually
exclusive) experiments without contradiction we may have to model
them by different sets of random variables and/or probability spaces
with different indecomposable elements $\omega$, $\omega'$ or
$\omega'',...$  and corresponding probability measures $P, P',
P'',...$ of the probability spaces $(\Omega, P)$, $(\Omega', P')$ or
$(\Omega'', P''),...$ respectively. This fact was well known to
Kolmogorov \cite{khr2}. {\it Bell criticized von Neumann
\cite{bellbook} for his dealings with incompatible experiments but
proceeded to put such experiments on one common probability space.
The existence of a common probability space, however, is not
guaranteed if the experiments are incompatible.}

A particular problem that arises in this connection stems from the
use of time $t$. For example, the time in the reference frame of the
measurement station that can be read on a clock does represent a
sense impression as defined by Mach and thus can be used in both
classical and quantum physics. However, time as such is not a random
variable and can certainly not be defined as a function on the
probability space that might be useful for a roulette. Measurement
times $t_1, t_2,..., t_J$ of a number $J$ of measurements can be
introduced, with some caution, as possible outcomes of a random
variable $T$ that is defined on some suitable probability space
\cite{hpfound}. The conventional inclusion of time in probability
theory, however, is to introduce time-labels for random variables.
This is done in the definition of stochastic processes. Before we
define a stochastic process we note that a given application in
science is not automatically guaranteed to have properties that can
be described as a stochastic process.

A stochastic process $A(t)$ can be regarded as a map \cite{brpe}:
\begin{equation}
A : \Omega \times T \rightarrow \Re^d \label{j08f1}
\end{equation}
for a fixed $t \in T$ \cite{brpe}. Here $\Re^d$ is a d-dimensional
vector with real components. In the cases considered below the
components assume only values of $\pm 1$ and $d \leq 4$. Then,
subsets of the sample space $B_1, B_2,..., B_J$ in $\Re^d$
corresponding to the discrete times $t_1, t_2,..., t_J$ are
introduced to define a joint probability
\begin{equation}
P(B_1, t_1; B_2, t_2;...; B_J, t_J) \equiv \mu(A(t_1) \in B_1,
A(t_2) \in B_2,A(t_J) \in B_J) \label{j08f2}
\end{equation}
that the process $A(t)$ takes on some value in $B_1$ at $t_1$, $B_2$
at $t_2$,..., and some value $B_J$ at $t_J$. The $A(t_1),
A(t_2),..., A(t_J)$ are by definition all (vector) random variables
on one common probability space. Because of this definition, each
stochastic process gives rise to the above family of joint
probabilities that must satisfy the 4 Kolmogorov consistency
conditions of which we only note the following two:
\begin{equation}
P( \Re^d, t) = 1 \label{j08f3}
\end{equation}
stating that the probability of the sure event is normalized to one.
Furthermore,
\begin{equation}
P(B_1, t_1; B_2, t_2;...; B_J, t_J) \geq 0 \label{j08f4}
\end{equation}
and refer the reader to \cite{brpe} for the other two.

It is important to note that the combination of two or more scalar
stochastic processes that are related to incompatible experiments to
form a vector stochastic process is not guaranteed to be possible
without contradictions. Assume, for example, that one stochastic
process, say $A({\bf a}, t)$ depends also on a certain setting $\bf
a$ and another $A({\bf b}, t)$ on a setting $\bf b$ and assume that
these settings are mutually exclusive at the same discrete time,
then the combined vector valued \cite{brpe} stochastic process
$(A({\bf a}, t),A({\bf b}, t))$ can not be defined for the same
sequence $t_1, t_2,..., t_J$. The corresponding probabilities of
Eq.(\ref{j08f4}) would all correspond to impossible events and their
sum is equal to zero which contradicts Eq.(\ref{j08f3}) (see section
3.2). For any particular science application of the concept of
stochastic processes, and particularly vector valued stochastic
processes, it is important to realize that consistency of the joint
probabilities is not guaranteed but {\it a premise} for Kolmogorov's
proof of the existence of sets of random variables on one
probability space \cite{khr2}, \cite{hpleiden}. The functions of any
particular science application can therefore not automatically be
treated as random variables.

\section{Bell's Theorem,
 Kolmogorov Consistency, Quantum Physics}

The author has attempted to find in the literature a formulation of
Bell's Theorem that proceeds in a scientific way by stating the
precise relations to the actual experiments from the view of both
classical and quantum probability as well as the precise
mathematical and physical premises and proceeding from these to
derive a demonstrable truth. The author found, in the original work
of Bell and in most subsequent work, the following sequence. Work is
started with the hypothesis of Einstein-Podolsky-Rosen that
``elements of physical reality...found by an appeal to results of
experiments and measurements" \cite{EPR} exist and are
mathematically represented by Einstein-local hidden variables.

In a first important step (i) Bell tries to show that the EPR theory
leads to a description of the EPR experiments by a number ($\geq 3$)
of random variables on one common probability space. This, in turn,
implies the existence and consistency of the higher order (2, 3 and
higher) joint probabilities that these random variables assume
certain values. It is this step that is criticized here by the
demonstration that for mutually exclusive (incompatible) experiments
definition of the corresponding functions on one common probability
space can not always be guaranteed and leads to logical
contradictions even for Einstein local classical physics. EPR stated
explicitly (second to last paragraph) that their {\it elements of
reality do not require} simultaneous measurability. If they are not
simultaneously measurable, they are not necessarily describable on
one common probability space because no $\omega^{act}$ needs to
exist that ``crystallizes" \cite{willi} them all into existence.
Incompatible experiments require, in general, the introduction of a
separate and different probability space. Their results can not
necessarily be contained in one common set of possible outcomes
without logical contradictions. This in turn means that neither the
existence of all the higher order distributions nor the existence of
more than two random variables on one common probability space are
guaranteed for EPR experiments.

In a second step (ii) the Bell inequalities are derived from the
existence of the higher order (2, 3 and higher) distributions. This
step is undoubtedly correct although it had been discussed before
Bell with its precise mathematical conditions by Vorob'ev
\cite{vorob}. Step (i) and (ii) lead then to the third and final
step (iii) the ``impossibility proof" showing that Bell's
inequalities are incompatible with quantum mechanical results and
experimental results of quantum optics for certain combinations of
pair expectation values (each corresponding to order 2 joint
probabilities). Then it is usually stated (see e.g. \cite{march})
``hence no local hidden-variable theory can reproduce exactly the
quantum mechanical statistical predictions. This is Bell's theorem".

We note, however, that if step (i) can not be proven from the basic
EPR hypothesis without additional unjustified and unjustifiable
assumptions, then the Bell inequalities do not follow from EPR and
the incompatibility of the Bell inequalities with quantum
correlations for certain setting combinations has no direct
significance with respect to physical reality or Einstein locality.

The importance of the first step (i) is explicitly stated by Mermin
\cite{mermin}, \cite{mermin1} and others \cite{legpriv0}. For
example, Mermin \cite{mermin} writes that the ``very compelling
Einstein-Podolsky-Rosen (EPR) argument would appear to {\it require}
the existence of just those higher-order distributions...", thus
Mermin implies that the mere hypothesis of the existence of elements
of physical reality, or his ``predetermined values" of possible
measurement outcomes, plus Einstein locality, does logically lead to
the existence and consistency of all higher order (2, 3 and higher)
joint probabilities. We denote this requirement of higher order (2,
3 and higher) consistency by RHOC. Leggett and Garg \cite{garg} even
assert that ``It immediately follows from (A1) that...we can
define...joint probability densities..." where (A1) refers to
``macroscopic realism" and the joint probability densities include
higher order ($2, 3...$). The author can not help suspecting that
Mermin, Leggett and Garg as well as many others (see e.g.
\cite{takagi}) did not realize the following: Kolmogorov's existence
proof shows only that {\it if} the higher order probability
distributions are consistent then there exist random variables on a
common probability space that reproduce these distributions and
neither proves the existence of random variables nor the existence
and consistency of all the higher order distributions for any given
application. The usual preamble in mathematical work ``Let $(\Omega,
F, P)$ be a Kolmogorov probability space" \cite{khr2} may indeed
have induced some to think that this can be automatically done for
any set of experiments and that $(\Omega, F, P)$ can be used as the
common domain for any number of functions even if incompatible or
mutually exclusive experiments and measurements are involved. We
will see that this is not the case.

As we will show below, Bell's mathematical formulation (BMF)
\cite{bell} and his parameter $\lambda$ neither prove the existence
of nor the necessity to use three or more functions on one common
probability space to describe EPR experiments. Such necessity, in
contrast to the mere existence of elements of reality, would indeed
impose certain limitations on the probabilities that (2, 3, 4...) or
more random variables assume certain values. This was shown already
before Bell by Vorob'ev \cite{vorob} and others.

We will show by general considerations and reference to classically
conceived counterexamples in this section (and by other methods in
section 4) that the arguments that have led Bell and his followers
to step (i) lead to logical contradictions for mutually exclusive
experiments and measurements and that Bell's definition of all
functions on one probability space is not a necessary requirement
even for problems of classical physics. Furthermore it will be
discussed in section 3.3 that step (ii) of Bell represents a special
case of the earlier classical mathematical work of Vorob'ev and
others that, ironically, also provides classical counterexamples for
the first step (i).

Before we show all of this we emphasize that no actual set of
experimental results (such as those of Aspect \cite{Aspect}) can
relate directly to all those higher order probability distributions
because the corresponding EPR experiments (with two settings on each
side) are incompatible and mutually exclusive at a given measurement
time or for a given entangled pair. Thus Bell and followers did not
and could not possibly have derived the existence of some of the
higher order ($\geq 3$) probability distributions by any direct
appeal to results of actual experiments. The experimental evidence
and the quantum theory for entangled pairs relate only to pair
correlations and thus to pairs of random variables and not to the
joint probabilities that three or more random variables assume
certain values.

\subsection{Bell's $\lambda$; Bell's basic assumptions}

We now turn to Bell's mathematical definitions and assumptions
\cite{bell} for spin related EPR experiments that involve entangled
pairs. As Bell, we consider spin $1/2$ particles in the singlet
state. We use for now Bell's notation $A({\bf a}, (\cdot)) = \pm 1$
in one wing and $B({\bf b}, (\cdot)) = \pm 1$ in the second wing of
the EPR-experiment \cite{bell}. Here $(\cdot)$ indicates additional
functional dependencies to be described below and $B({\bf b},
(\cdot)) = -A({\bf b}, (\cdot))$ . The bold faced subscripts $\bf a$
and $\bf b$ are unit vectors of three dimensional space and describe
the measurement settings of the Stern Gerlach magnets. The
corresponding quantum operators are the spin matrices $\sigma_{\bf
a}$ and $\sigma_{\bf b}$ \cite{hpwqi}.

Bell also has introduced a parameter $\lambda$ that characterizes
``elements of reality" related to the entangled pair. Bell noted:
``It is a matter of indifference in the following whether $\lambda$
denotes a single variable or a set or even a set of functions, and
whether the variables are discrete or continuous". This has given
Bell's followers the impression that $\lambda$ can be ``anything".
Bell's use of $\lambda$ implies, however, important restrictions
because he assumes that $\lambda$ is related to the entangled pair
emanating from a single source and that the possible spin outcomes
are functions of $\lambda$. The spin outcomes considered by Bell and
his followers are usually denoted by $A({\bf a}, \lambda), A({\bf
b}, \lambda), A({\bf d}, \lambda)$ in station $S_1$ and $B({\bf b},
\lambda), B({\bf c}, \lambda)$ in station $S_2$ respectively with $B
= - A$ for equal settings. Thus, $\lambda$ represents a ``package"
sent from a source and the set of all packages forms the domain of
all functions.

Bell's next step, after introducing the parameter $\lambda$ and
functions $A, B$ of $\lambda$, is the {\it assumption} of a
probability distribution $\rho(\lambda)$ with
\begin{equation}
\int \rho(\lambda) d \lambda = 1 \label{n08j9}
\end{equation}
This means that all functions are defined on the common sample space
constituted by the $\lambda$'s and because of Eq.(\ref{n08j9}) are
defined one one common probability space. Bell thus describes the
spin-outcomes of incompatible (mutually exclusive) experiments and
measurements by functions:
\begin{equation}
A({\bf a}, \lambda), A({\bf b}, \lambda), B({\bf b}, \lambda), B({\bf
c}, \lambda) = \pm 1\label{f11}
\end{equation}
that are all defined on the same domain and are algebraically
manipulated accordingly. Furthermore, Bell emphasizes, that because
of Einstein locality $\lambda$ is independent of the settings. This
is indeed guaranteed by the delayed choice experiments implemented
by Aspect and others if $\lambda$ describes entities emanating from
a source. The above represents the essence of Bell's mathematical
formulation (BMF) and we summarize the important assertions:
($\alpha$) $\lambda$ may represent any Einstein local set of
elements of physical reality, ($\beta$) $\lambda$ is independent of
the instrument settings and the equipment is fully characterized by
a setting vector and ($\gamma$) all functions describing spin
outcomes (three or more) are defined on one common probability space
even though they describe incompatible experiments. We discuss below
examples of classical physics that contradict the first two of these
assertions and show that the third is an assumption that can not be
proven.

\subsection{Counterexamples for BMF, RHOC}

\subsubsection{Generality of $\lambda$}

The following counterexample is designed to show that the assumption
of a general $\lambda$ (assertion $(\alpha)$) leads to a
contradiction when step (i) is attempted for mutually exclusive
experiments. Bell stated that  ``...$\lambda$ denotes...a single
variable or a set...discrete or continuous..." The task in step (i)
of the Bell theorem is then to prove that the functions of
Eq.(\ref{f11}) that describe the spin outcomes and are defined on
sets of arbitrary general Einstein-local elements of reality as
defined by EPR (and symbolized by $\lambda$) have consistent higher
order joint distributions as for example:
\begin{equation}
P(\lambda: A({\bf a}, \lambda) = + 1, A({\bf b}, \lambda) = + 1, B({\bf
b}, \lambda) = - 1, B({\bf c}, \lambda) = + 1) . \label{f41}
\end{equation}

According to Bell's statement above we may substitute for $\lambda$
the set $\omega, t_j$ where $\omega$ is an element of a probability
space $(\Omega, P)$ and $t_j$ the clock time of measurement in the
frame of the measurement stations. The clock time $t_j$ is an
element of reality even in the sense of Mach and therefore a
fortiori in the sense of EPR, and $\omega$ can correspond to any
element of reality (also the sense of Mach) drawn from an urn or
generated by a computer. Bell's functions form then a classical
vector valued stochastic process involving vectors:
\begin{equation}
(A({\bf a}, \omega, t), A({\bf b}, \omega, t), B({\bf b}, \omega,
t), B({\bf c}, \omega, t)) \label{f12}
\end{equation}
The first term of the probability distribution given in
Eq.(\ref{j08f2}) and the corresponding measure $\mu$ is:
\begin{equation}
\mu (A({\bf a}, \omega, t_1) \in B1, A({\bf b}, \omega, t_1) \in
B_1, B({\bf b}, \omega, t_1) \in B_1, B({\bf c}, \omega, t_1) \in
B_1 ...) \label{f42}
\end{equation}
If we take now into account that Bell's functions describe mutually
exclusive experiments and measurements, then we encounter a serious
logical hurdle with BMF. Because the event of outcomes with
different settings in the same station to happen at the same time
$t_1$ is the impossible event and the same is true for all the other
measurement times, all the corresponding probabilities are zero and
we therefore have as mentioned before and proven in \cite{hpfound}
in great detail the result zero for all the joint probabilities.
This, however, is in contradiction to the second consistency
requirement of Eq.(\ref{j08f3}). Therefore the higher order joint
probabilities do not fulfill the consistency conditions and $(A({\bf
a}, \omega, t), A({\bf b}, \omega, t), B({\bf b}, \omega, t), B({\bf
c}, \omega, t))$ can not be the vector of a multivariate stochastic
process.

This disproves the assertion that the mere existence of elements of
reality, or of corresponding predetermined values, or of ``reality
at the macroscopic level" \cite{garg}, \cite{takagi} and use of
Einstein locality {\it requires} (RHOC) the existence and
consistency of those higher order distributions. Another similar
counterexample treating the measurement times as values that random
variables may assume has been presented with even greater
mathematical rigor in \cite{hpfound}. Furthermore, Vorob'ev
\cite{vorob} gives an example from the theory of games that can also
be used as a counterexample to RHOC and BMF.

The above example demonstrates that the settings of one station
(e.g. $\bf a$, $\bf b$) and the measurement times taken as elements
in the sense of Mach can not be varied independently. If we change
the setting we must permit a change in the stations clock-time to
occur. For this reason we have used in the past a slightly different
notation with the settings as a subscript to make it clear that a
different setting implies the use of a different function that
represents the spin measurements. We also change the notation in
this way from now on:
\begin{equation}
A({\bf a}, ( \cdot )) \rightarrow A_{\bf a}(( \cdot )) \label{march07n1}
\end{equation}
with similar provisions for all other settings and the functions $B$.

In the next two sections we show that the assumption of one common
probability space is, from the viewpoint of applications including
classical and quantum physics a big assumption indeed and can in
general not be justified. It will also be shown that the very
introduction of time $t$ gives a proof for the lack of generality of
Bell's function space, a proof that uses only elements of reality
that all can be encountered in actual EPR experiments and that is
also fully consistent with Einstein's relativity.

\subsubsection{Instrument parameters $\lambda_{\bf a}(t, s_1)$}

The following represents a counterexample of assertion $(\beta)$.
The introduction of time $t$ into our reasoning permits the
construction of local instrument parameters that are (see end of
section 4) not stochastically independent. Einstein-local time and
setting dependent instrument variables may exist and reflect the
physics, including unknown physical effects, of the respective
measurement instruments. We have described such instrument
parameters in detail in \cite{hppnas}, \cite{hpfound} and denote
them here by $\lambda_{\bf a}(t, s_1), \lambda_{\bf b}(t, s_1),
\lambda_{\bf d}(t, s_1)$ for station $S_1$ and $\lambda_{\bf b}(t,
s_2), \lambda_{\bf c}(t, s_2)...$ for station $S_2$ respectively
where $t$ stands for discrete measurement times shown e.g. by clocks
in the respective stations. Here $s_1, s_2$ denote some additional
parameters characteristic for the backward light-cone of the
measurement in the respective station. Now we can let the functions
$A, B$ depend on these additional parameters e.g. $A = A_{\bf a}(
\lambda, \lambda_{\bf a}(t, s_1))$. Thus the functions $A, B$ are
now functions of one more variable that does depend on the setting
and may be defined on a sample space different from that of Bell's
$\lambda$ and therefore also on a different probability space say,
$\Omega', P'$. In principle, we may also introduce different sample
spaces and different probability spaces for each given fixed
measurement time as will be discussed immediately. When we consider
all the different incompatible measurements, all the different
settings, measurement times and possible sample spaces then we can
see the enormous over-simplification that Bell made even from a
viewpoint of classical physics. We will discuss this more explicitly
in the next section and just add here two remarks.

Because of the importance of these station parameters we quote
Marchildon \cite{marchleid} who has described their physical
interpretation succinctly: ``Firstly, and in the spirit of quantum mechanics,
neither particle has a precise value of any of its spin components
before measurement. Rather, the particles and the instruments
jointly possess information that is sufficient for deterministic
values to obtain upon measurement. Secondly, the dependence of the
instrument's random variables on some universal time allows for a
stochastic dependence of measurement results on one another...".

Finally we note that Mermin \cite{mermin1} has considered our
suggestion of instrument variables but immediately has dropped the
setting index to replace:
\begin{equation}
A_{\bf a}( \lambda, \lambda_{\bf a}(t, s_1)) \rightarrow A_{\bf a}(
\lambda, \lambda(t, s_1)) \label{ja311}
\end{equation}
This is just one of several inaccuracies that Mermin introduced. We
can see the problem by considering that in his notation:
\begin{equation}
A_{\bf b}( \lambda, \lambda_{\bf b}(t, s_1)) \rightarrow A_{\bf b}(
\lambda, \lambda(t, s_1)) \label{ja312}
\end{equation}
which now uses the same $\lambda(t, s_1)$ for the two different
variables $\lambda_{\bf a}(t, s_1)$ and $\lambda_{\bf b}(t, s_1)$.
We will show in section 4 a very convincing example that the
dropping of indices that label different entities is unbecoming.

The additional instrument parameters that are, in general, different
for different settings and measurement times call also for different
probability spaces that are necessary to describe them.
Considerations of generality require in fact different probability
spaces for each setting pair and each measurement time. This is
explained in detail next.

\subsubsection{The general function space that Bell should have
admitted}

Consider now the functions $A, B$ and let us use the notation that
we have introduced for stochastic processes, noting that we do {\it
not} assume here that these functions indeed are the random
variables of a stochastic process and therefore defined on a single
probability space. Because the single measurements in station $S_2$
are made in essence ``simultaneously" with those of station $S_1$
for each entangled pair, we can use the same probability space for
the random variables $B$ and $A$ for one given setting on each side.
For any single experiment we can use the same measurement time $t_j$
for both $A$ and $B$, or also a different measurement time $t'_j$
for $B$ just corresponding to the same entangled pair if we wish to
do so. We will in the following, for simplicity in the notation
assume the same measurement time $t_j$ for a given entangled pair
but note again that our notation and everything we claim can easily
accommodate a different time $t'_j$. The subscript of the
measurement times is the same for a given entangled pair. We need to
make sure, however, that we use for different setting pairs
different time sequences. Note that we dropped for simplicity the
parameters $s_1, s_2$ that need to be included, in general, along
with $t_j$ and $t_j'$ respectively. For setting $\bf a$ in station
$S_1$ and $\bf b$ in station $S_2$ we use then functions $(A_{\bf
a}({\omega^1},t_{1}), A_{\bf a}({\omega^2},t_{2}),..., A_{\bf
a}({\omega^J},t_{J}))$ and $(B_{\bf b}({\omega^1},t_{1}), B_{\bf
b}({\omega^2},t_{2}),..., B_{\bf b}({\omega^J},t_{J}))$
respectively. We introduce additional functions $(A_{\bf
b}({\omega^{J+1}},t_{J+1}), A_{\bf
b}({\omega^{J+2}},t_{J+2}),...,A_{\bf b}({\omega^{2J}},t_{2J}))$ and
corresponding functions $(B_{\bf c}({\omega^{J+1}},t_{J+1}), B_{\bf
c}({\omega^{J+2}},t_{J+2}),...,B_{\bf c}({\omega^{2J}},t_{2J}))$. We
also need to introduce another function set $(A_{\bf
a}({\omega^{2J+1}},t_{2J+1}), A_{\bf
a}({\omega^{2J+2}},t_{2J+2}),...,A_{\bf a}({\omega^{3J}},t_{3J}))$
and, to complete the correspondence to Bell, the set $(B_{\bf
c}({\omega^{2J+1}},t_{2J+1}), B_{\bf
c}({\omega^{2J+2}},t_{2J+2}),...,B_{\bf c}({\omega^{3J}},t_{3J}))$.
In this way we have generated a very general system of functions on
very general domains. Note that we admit $3J$ different probability
spaces $(\Omega^1, P^1), (\Omega^2, P^2)..., (\Omega^{3J}, P^{3J})$.
This means that we use the elements of a different sample space for
each different time index (and, in general, also for each different
$s_1$ and $s_2$). We call this general set of functions and
probability spaces the generalized EPR set and denote it by GEPRS.
The question whether any subset of the GEPRS set or of products such
as $A_{\bf a}(\omega^j, t_{j}) B_{\bf b}(\omega^j, t_{j})$ with $j =
1,...,J$ or $A_{\bf a}(\omega^i, t_{i}) B_{\bf b}(\omega^i, t_{i})$
with $i = J+1, ,2J$ etc. can be defined on a common probability
space and therefore forms a stochastic process is, as we saw above
and will further demonstrate below, a question that has no trivial
answer and needs to be checked for any particular application. It is
certain (see next section) that the full GEPRS set can not always be
defined on one common probability space and can therefore not always
be regarded to form a (vector) stochastic process.

\subsection{Bell, Vorob'ev, quantum mechanics, cyclicity}

This section is not entirely self contained and the reader is
referred to the detailed proofs of \cite{hpfound}. We just wish to
emphasize the following. The contradiction that Bell obtained is not
rooted in the specific quantum mechanical pair expectation values
and their experimental confirmation but in the additional
requirement that three or more functions are defined on the same
common probability space even though they describe incompatible
experiments. This additional requirement can only be met if the
higher order distributions such as the probability that the three
(or more) random variables assume certain values are consistent with
the joint pair distributions. The existence and consistency of these
higher order ($\geq 3$) distributions is not required by quantum
mechanics that does not deal with random variables. As we have seen,
it is also not required and can not be met for certain classical
vector valued stochastic processes that involve mutually exclusive
experiments. EPR have not required the existence and consistency of
the higher order ($\geq 3$) distributions because they did expressly
consider elements of physical reality that can not be measured
simultaneously. According to Vorob'ev's work, the additional
requirement results in mathematical constraints even for sets of
pair probability distributions if they exhibit a
topological-combinatorial ``cyclicity" \cite{vorob}.

It will be shown below by using Vorob'ev's classical line of
reasoning that, if we consider the functions of section 3.2, we may
concatenate the $3J$ admissible different probability spaces for the
$3J$ function pairs to form 3 mutually exclusive function pairs with
each different pair defined on a different abstract probability
space. However, a contradiction may arise if one attempts to further
concatenate the pairs corresponding to mutually exclusive
experiments and to describe them on a single common probability
space. We first prove that no contradiction arises if compatible
(i.e. not mutually exclusive) EPR experiments are concatenated on
one probability space. We use, as Bell did, the relation $B_{\bf a}(
\omega) = - A_{\bf a}( \omega)$ just for the sake of an efficient
presentation.

{\bf Lemma:} For one given setting in each of the two stations, and
thus for a given set of compatible experiments, it is possible to
disregard the different measurement times ($t_j$, $t_i$, etc.) of the previous
section, and to find a single abstract probability space on which the functions
$A, B = \pm 1$ of GEPRS can be defined while still obtaining the pair
expectation value prescribed by quantum mechanics.

{\bf Proof:}

The goal is to obtain the pair expectation value $M(A_{\bf a}(
\omega)B_{\bf b}( \omega)) = -M(A_{\bf a}( \omega)A_{\bf b}(
\omega))$ prescribed by quantum mechanics \cite{bell}, \cite{march}:
\begin{equation}
M(A_{\bf a}( \omega)A_{\bf b}( \omega)) = -\langle
\psi_B|\sigma_{\bf a}\otimes\sigma_{\bf b}| \psi_B \rangle = {\bf a}
\cdot{\bf b} \label{aug071}
\end{equation}
where $\otimes$ denotes the tensor product and $|\psi_B \rangle$ is
the Bell wave function:
\begin{equation}
|\psi_B>=\frac{1}{\sqrt{2}}\left(\left(\begin{matrix}1 \\
0\end{matrix}\right)\otimes \left(\begin{matrix}0 \\
1\end{matrix}\right)-\left(\begin{matrix}0 \\ 1\end{matrix}\right)
\otimes\left(\begin{matrix}1 \\
0\end{matrix}\right)\right).
\end{equation}
The following joint probability measure results in $M(A_{\bf a}(
\omega)A_{\bf b}( \omega))$ of Eq.(\ref{aug071}) as can be found by
inspection (see \cite{hpwqi}, \cite{hpleiden}):
\begin{equation}
P(\omega: A_{\bf a}( \omega) = (-1)^n, A_{\bf b}( \omega) = (-1)^k )
= {\frac{1} {4}}(1+ (-1)^{n+k} {\bf a} \cdot {\bf b}) \label{aug072}
\end{equation}
Here $k, n = 1, 2$. The probability measure so defined also fulfills
$M(A_{\bf a}(\omega)) = M(A_{\bf b}(\omega)) = 0$ as required by
quantum mechanics \cite{hpfound}. Thus, EPR spin experiments as
discussed by Bell and restricted to precisely one setting on each
side, i.e. to compatible experiments, can be described by one
abstract probability space with elements $\omega \in \Omega$ that
represent both source and equipment variables. We have for all
probabilities $0 \leq P \leq 1$ and no contradiction of the Bell
type occurs. We have also shown previously that the above random
variables and abstract probability space can be simulated by use of
a classical computer \cite{hpwqi} and have also given an elaborate
Einstein-local mathematical model that includes the setting and time
dependent equipment parameters \cite{hppnas1}.

Before we proceed, we recall that the elements of the GEPRS set are
functions on numerous domains that can at least, in principle, not
necessarily be defined on one probability space. To single out the
setting pair as the only indication for differences in the functions
is an arbitrary choice such as the choice of the color of a shirt to
characterize a person. It is therefore not a trivial fact that all
the possible domains can be replaced by one common probability
space. It is also not trivial that, as shown in the lemma, one can
disregard the functional dependencies on the measurement times
($t_j, t_i$ etc.) and still recover some of the major results of
quantum physics. We will see, however, that this is not always
possible when incompatible experiments are involved. We show also
that the ``impossibility" that Bell actually found does not arise
from the mere fact that we wish to obtain certain pair expectation
values such as ${\bf a} \cdot {\bf b}$ but necessitates the
additional requirements that three or more functions describing
incompatible composite (multi stage) experiments are defined on a
single common probability space. These facts follow from the
detailed proof of the following theorem.

{\bf Theorem 1:}

Consider functions $A_{\bf a}, A_{\bf b}, B_{\bf b}, B_{\bf c} = \pm
1$ with $A_{\bf b} = -B_{\bf b}$ etc.. Consider further three or
more pair products of these functions and all expectation values
that can be obtained for these pair products by use of a different
probability space for each different pair. It is then in general
impossible to find a single common probability space $(\Omega, P)$
that reproduces all these expectation values for the pair products.
The reason for this impossibility can always be traced (also for
functions with a more general range) to a combinatorial-topological
cyclicity inherent in the set of the pair products and defined by
Vorob'ev \cite{vorob}. Expectation values from quantum mechanics
such as ${\bf a} \cdot {\bf b}$, ${\bf a} \cdot {\bf c}$ and ${\bf
b} \cdot {\bf c}$ corresponding to the cyclical products $A_{\bf
a}A_{\bf b}, A_{\bf a}A_{\bf c}$ and $A_{\bf b}A_{\bf c}$
respectively can, for certain settings ${\bf a}, {\bf b}, {\bf c}$,
serve as an example.

{\bf Proof:}

We present only an outline of the major ideas of the proof with
references to previous work that contain the precise and complete
elements of the proof as well as the complete solution of the
consistency problem.

It was shown in \cite{hpfound} that Bell's inequalities represent a
special case of the more general mathematical framework of Vorob'ev
\cite{vorob}who showed that {\it "..it is not always possible to
construct a vector random variable with given consistent
projections."} Vorob'ev's \cite{vorob} work gives precise
mathematical conditions for the validity of his statements and
theorems and the serious reader should at least understand Theorem 1
of \cite{hpfound} and the first page of \cite{vorob}. However, the
essence is this:

{\it The pairs of random variables $A_{\bf a}( \omega)A_{\bf b}(
\omega)$, $A_{\bf a}( \omega)A_{\bf c}( \omega)$ and $A_{\bf b}(
\omega)A_{\bf c}( \omega)$ form a ``closed loop" or display a
``cyclic behavior" \cite{vorob}. Then, once the pair distributions
of $A_{\bf a}( \omega)A_{\bf b}( \omega)$ and $A_{\bf a}(
\omega)A_{\bf c}( \omega)$ are given one can not choose that of
$A_{\bf b}(\omega)A_{\bf c}( \omega)$ with complete freedom and at
the same time require that $A_{\bf a}( \omega), A_{\bf b}(
\omega),A_{\bf c}(\omega)$ are all random variables defined on one
common probability space.}

In terms of the algebra of random variables one finds the well known
constraints on the possible outcomes for four setting pairs (see
section 4 for details):
\begin{equation}
\Gamma = A_{\bf a}({\omega})B_{\bf b}({\omega}) +
A_{\bf a}({\omega})B_{\bf c}({\omega}) +
A_{\bf d}({\omega})B_{\bf b}({\omega})
-A_{\bf d}({\omega})B_{\bf c}({\omega}) =
\pm 2 \label{march3n1}
\end{equation}
which leads to the Bell-type inequality $\Gamma \leq 2$. Thus the
range (codomain) of the function $\Gamma$, which plays a very
significant role in the framework of Bell, is restricted to $\pm 2$
because of the assumption of a single domain for the cyclically
arranged functions. Without the requirement of one domain, that has
no basis for incompatible experiments, the range of $\Gamma$ would
be $\pm 4$. The use of the pair expectation values suggested by
Eq.(\ref{aug071}) is possible according to the lemma for any single
setting pair and the involved functions $A_{\bf a}( \omega)A_{\bf
b}( \omega)$ can be defined on one probability space. We can also
obtain the quantum result for the pair expectation of $A_{\bf a}(
\omega')A_{\bf c}( \omega')$ by using a different probability space
$(\Omega', P')$ etc.. However we can not require for any physical or
mathematical reason that all three (or four) functions $A_{\bf a}(
(\cdot)), A_{\bf b}( (\cdot)),A_{\bf c}( (\cdot))$ etc. must always
be defined on one common probability space. If they are not, then
Eq.(\ref{march3n1}) makes no mathematical sense. The constraint on
the range of $\Gamma$ is therefore not required by the physics of
the problem but only by the desire to use more than two random
variables and one domain (probability space) to model that physics.

In summary we have found that the existence and consistency of all
higher order distributions has been correctly disproved by Bell in
his step (iii) for cyclically arranged functions describing
incompatible spin related EPR experiments. However, the existence
and consistency of third order joint probabilities is neither
required by experiment nor by quantum theory and was only assumed by
Bell without justification. Furthermore, Bell also could have
already disproved the general validity of step (i) by using the
example of a classical vector valued stochastic process or
Vorob'ev's classical reasoning and his example from the theory of
games. But then Bell would have disproved altogether that step (i)
must follow from the mere hypothesis of existing elements of reality
and Einstein locality. If step (i) is disproved in that way, then
the third step (iii) looses its connection to EPR. Because the
higher order ($\geq 3$) distributions have also no connection to
actual experiments, Bell's inequalities have therefore the following
meaning. Composite functions such as $\Gamma$ that are created by a
cyclical arrangement of other functions exhibit a constraint on
their possible range (co-domain). Certain ranges can not be achieved
by use of one domain only. Incompatible experiments may have to be
defined by their very nature on different domains and $\Gamma$ may
therefore not lend itself to describe incompatible experiments by
use of one domain without contradiction.

\section{Variations on
Clauser-Horn-Shimony-Holt (CHSH) by Mermin, Leggett and Peres }

The Clauser-Horn-Shimony-Holt \cite{chsh} no-go proof has been
discussed extensively in the literature. While their original work
still contains the parameter $\lambda$ and uses Bell's assumptions
that imply one common probability space, textbooks such as
\cite{peres} and \cite{leggett} present proofs that do not include
any explicit reference to $\lambda$. In spite of rigorous
mathematical discussions in the literature \cite{khrbook}, it is
therefore often believed that at least these proofs require ``only
the existence of predetermined values...independent of the local
mechanism that produces them" \cite{mermin1} and thus show that the
EPR hypothesis by itself results in a constraint on the collection
of actual outcomes. Some even believe that these proofs relate
directly to experimental data and represent therefore a proof of the
area of statistics. We show in this section (theorem 2) that the
expression for the hypothetical predetermined values that is used in
the texts by Leggett \cite{leggett}, Peres \cite{peres} and others
as well as in the publications of Mermin \cite{mermin},
\cite{mermin1} does not relate directly to the actual experimental
data and therefore can also not be used to construct the sample
space of Kolmogorov's probability theory. Any scientifically serious
pre-statistics (probability theory) needs a sample space related in
a clear and direct way to the actual experiments. It follows then,
that CHSH and the variations of Mermin, Leggett and Peres are either
just a reformulation of Bell's work who attempted to describe the
actual experiments by using functions defined on one common
probability space, or follow from the unjustifiable assumption of
conditional stochastic independence as discussed at the end of this
section.

To show all of this we start from the actual experimental results.
We use the notation for spin $1/2$ related experiments in order to
relate directly to Bell's work. However, all that is claimed and
used here with respect to the experiments is also valid for actual
optical experiments, such as the celebrated Aspect experiment.
Consider then experiments taken with equal probability of $P = 1/4$
for magnet (or Kerr cell etc.) setting pairs $({\bf a}, {\bf b})$,
$({\bf a}, {\bf c})$, $({\bf d}, {\bf b})$ and $({\bf d}, {\bf c})$
respectively as proposed by CHSH. We denote the actual experimental
outcomes by the lower case symbols $a_{\bf a}, a_{\bf d}$ in one
wing of the experiment and $b_{\bf b}, b_{\bf c}$ in the other with
the subscripts indicating the magnet settings in the respective
stations. We also add integer indices $j = 1,...,J; i = J+1,...,2J;
l=2J+1,...,3J; s = 3J+1,...,4J$ that enumerate the collected
experimental data using $j$ for setting pairs ${\bf a}, {\bf b}$,
$i$ for ${\bf a}, {\bf c}$, $l$ for ${\bf d}, {\bf b}$ and $s$ for
${\bf d}, {\bf c}$ respectively; J being just a large number. In
this way we can represent all experimental sequences and their
addition/subtraction as performed by CHSH by using about $J$
expressions of the form:
\begin{equation}
\gamma_{exp}^{j,i,l,s} =: a_{\bf a}^j \cdot b_{\bf b}^j + a_{\bf
a}^i \cdot b_{\bf c}^i +  a_{\bf d}^l \cdot b_{\bf b}^l - a_{\bf
d}^s \cdot b_{\bf c}^s  .\label{n08j1}
\end{equation}
The equality of the superscript in the products of Eq.(\ref{n08j1})
pays attention to the fact that the experiments in the two wings
correspond to the same entangled pair. We have introduced, for the
sake of generality, different superscripts that indicate
different entangled pairs as well as the possibility of different
instrument variables.

It is important to note that $\gamma_{exp}^{j,i,l,s}$ can
assume the values:
\begin{equation}
\gamma_{exp}^{j,i,l,s} = 0, \pm 2, \pm 4 \label{nnn08j1}
\end{equation}
This follows from the fact that the actual experimental results can
only assume values of $\pm 1$ for each of the factors of
Eq.(\ref{n08j1}). The expectation value $M(\gamma_{exp}^{j,i,l,s})$
of $\gamma_{exp}^{j,i,l,s}$ for the $J$ quadruple experiments can
then be calculated from:
\begin{equation}
M(\gamma_{exp}^{j,i,l,s}) = \frac {1} {J}[\sum_{j=1}^J a_{\bf a}^j
\cdot b_{\bf b}^j + \sum_{i=J+1}^{2J} a_{\bf a}^i \cdot b_{\bf c}^i
+ \sum_{l=2J+1}^{3J} a_{\bf d}^l \cdot b_{\bf b}^l
-\sum_{s=3J+1}^{4J} a_{\bf d}^s \cdot b_{\bf b}^s]  \label{nnn08j2}
\end{equation}

We know from actual experiments of the Aspect type that for a
specific choice of setting pairs we have:
\begin{equation}
M(\gamma_{exp}^{j,i,l,s}) = 2 + \delta \label{nnn08j3}
\end{equation}
where the experimentally found $\delta$ may be as much as about
$0.8$ (Mermin speaks about 40 \% above the value 2 \cite{mermin1})
which also agrees with quantum theory that results in a supremum of
$2 \sqrt{2}$ for M. We therefore can state the following theorem:

$\bf Theorem 2:$

The actual experimental results for the composite
$\gamma_{exp}^{j,i,l,s}$ that are obtained from $J$ experimental
quadruples and are used to calculate the expectation value $M$ by
using Eq.(\ref{nnn08j2}) and result in Eq.(\ref{nnn08j3}) must
contain about $R$ terms having values of $\gamma_{exp}^{j,i,l,s} =
4$ with $R \geq \frac {J \delta} {2}$.

$\bf Proof$ (by contradiction):

Assume that the results of $J$ (quadruple-composite) experiments
contain the value $\gamma_{exp}^{j,i,l,s} = 4$ only in numbers $R <
\frac {J \delta} {2}$. We know that we can only encounter the
experimental values of $\gamma_{exp}^{j,i,l,s} = 0, \pm 2, \pm 4$
and that therefore
\begin{equation}
M(\gamma_{exp}^{j,i,l,s}) = {\frac {1} {J}}(-4O -2P +2Q +4R)
\label{adjan251}
\end{equation}
where $O, P, Q, R$ are positive integers that indicate the
occurrences of the values $-4, -2, 2, 4$ respectively. We also may
have $S$ occurrences of the value zero and therefore have $O + P + Q
+ R + S = J$. From Eq.(\ref{adjan251}) and the assumption of $R <
\frac {J \delta} {2}$ we have:
\begin{equation}
M(\gamma_{exp}^{j,i,l,s}) \leq {\frac {1} {J}}(+2Q +4R) \leq 2 +
\frac {2R} {J} < 2 + \delta \label{adjan252}
\end{equation}
which contradicts the experimental result of Eq.(\ref{nnn08j3}) and
thus completes the proof.

Mermin, Leggett, Peres and others do not deal with the actual
experimental values of $\gamma_{exp}^{j,i,l,s}$ as defined above and
do not present a proof of the area of statistics. Instead they use a
very {\it different} quadruple $\gamma^{j}$ \cite{mermin1}:
\begin{equation}
\gamma^{j} = a_{\bf a}^j \cdot b_{\bf b}^j + a_{\bf a}^j \cdot
b_{\bf c}^j + a_{\bf d}^j \cdot b_{\bf b}^j - a_{\bf d}^j \cdot
b_{\bf c}^j \label{n08j4a}
\end{equation}
As one can easily see we have:
\begin{equation}
\gamma^{j} = \pm 2 \text{with $j=1,...,J$} \label{n08j5}
\end{equation}
$\gamma^{j}$ clearly has no relation to the actual experiments
because the experimental values represented by
$\gamma_{exp}^{j,i,l,s}$ must, according to theorem 2 also assume
$+4$ with statistical significance. Therefore $\gamma^{j}$ can also
not have anything to do with the sample space of Kolmogorov's
probability theory or a sample space of any equivalent reasonable
pre-statistics. To use $\gamma^{j} = \pm 2$ compares to the attempt to model
a game with a possible outcome of 5 different numbers by using only
two numbers in the mathematical model of the game. Or, in terms of
the even-odd pairs of section 2, one just can not explain the
experiment and leave out e.g. the combinations $eo, oo$.

Thus, the use of the quadruples $\gamma^{j} = \pm 2$ that are not
directly linked to experiments requires additional justification.
One needs to take a first step, corresponding to (i) in section 3,
that leads from the EPR hypothesis of elements of reality to the
necessity of using $\gamma^{j} = \pm 2$. Leggett and Peres use as a
basis a counterfactual argument that they believe to be justified
{\it if} the counterfactual EPR reasoning is taken to be justified.
We have discussed this extensively in \cite{hppnas}, \cite{hpfound}
and have shown that Leggett and Peres add an incorrect assumption to
EPR's valid counterfactual reasoning. We summarize these previous
findings below for completeness but first describe Mermin's
\cite{mermin1} equally false justification of using $\gamma^{j} =
\pm 2$.

Mermin writes ``We wish to test the postulate that the result
$a_{\bf x}(j) = \pm 1$...for measurements of the jth pair along any
direction $\bf x$ is predetermined...by hidden variables for all
possible $\bf x$..." where $\bf x$ stands for the possible settings
of Eq.(\ref{n08j4a}). Formulated like this, Mermin's ``postulate" is
not general and does, in general, not follow from the EPR hypothesis
because it links the index $j$ exclusively to the entangled pair
(just as Bell did with $\lambda$) and disregards possible elements
of reality related to the instruments and their settings. As we have
seen in section 3, a general approach must also admit setting
dependent instrument variables. We also know from section three that
the factors of $\gamma^{j}$ can not necessarily be regarded as
values that random variables defined on one probability space
assume. Disregarding all the problems related to the sample space of
a valid pre-statistics, Mermin proceeds to ``average uniformly over
all...pairs", $4J$ in our notation, to obtain an expectation value
of the following entity (that neither has any well defined
connection to the actual experiments nor to a corresponding {\it
general} set of EPR's elements of reality):
\begin{equation}
M(\gamma^{j}) = {\frac {1} {4J}}[ \sum_{j=1}^{4J} a_{\bf a}^j \cdot
b_{\bf b}^j + \sum_{j=1}^{4J} a_{\bf a}^j \cdot b_{\bf c}^j +
\sum_{j=1}^{4J} a_{\bf d}^j \cdot b_{\bf b}^j - \sum_{j=1}^{4J}
a_{\bf d}^j \cdot b_{\bf c}^j] \label{adjan253}
\end{equation}

Now Mermin claims that by ``standard sampling arguments" (a) each
sum in Eq.(\ref{adjan253}) will be very close to the corresponding
sum of Eq.(\ref{nnn08j2}) and (b) that at the same time
$M(\gamma^{j}) \leq 2$. Mermin does not give any literature
reference for this ``standard sampling argument" and only explains
that because of the random choice of setting the ``indices $j$...
constitute a random sample of the full set of indices..." which is
``surely what Asher (Peres), who has a well known distaste for being
explicit about what should be obvious \cite{mermin1}, had in
mind...". Because of the fact that in general the indices represent
not only the entangled pairs but also instrument variables, this
statement is false: the instrument variables can not be sampled
randomly. There are in Eq.(\ref{nnn08j2}) $4J$ entangled pairs but
only $J$ corresponding pairs of instrument variables that are
encountered for each of the 4 given setting pairs. These pairs of
instrument variables are, in general, different for different
setting pairs. The instrument variables $\lambda_{\bf a},
\lambda_{\bf b}$ may even have nothing at all in common with
$\lambda_{\bf d}, \lambda_{\bf c}$. Nevertheless, Mermin's claim (a)
that each sum in Eq.(\ref{nnn08j2}) will be very close to the
corresponding sum of Eq.(\ref{adjan253}) is true (under certain
mathematical conditions \cite{willi}) even if we involve instrument
variables. We know from the lemma in section 3, that the functions
of section 3.2 that involve only one pair of settings and therefore
correspond to compatible experiments can be concatenated on one
abstract probability space. In other words we can represent products
of e.g. the functions $(A_{\bf a}({\omega^1},t_{1}), A_{\bf
a}({\omega^2},t_{2}),..., A_{\bf a}({\omega^J},t_{J}))$ and $(B_{\bf
b}({\omega^1},t_{1}), B_{\bf b}({\omega^2},t_{2}),..., B_{\bf
b}({\omega^J},t_{J}))$ by products $A_{\bf a}({\omega_{abs}})B_{\bf
b}({\omega_{abs}})$. We have added here the subscript $abs$ to
emphasize that we are dealing with an abstract probability space
that arises from a concatenation of many sample and probability
spaces that in turn are related to the different elements of reality
in the instruments. For the functions involving the other setting
pairs we have then, in general different sample/probability spaces
e.g. $A_{\bf a}({\omega_{abs}'})B_{\bf c}({\omega_{abs}'})$, $A_{\bf
d}({\omega_{abs}''})B_{\bf b}({\omega_{abs}''})$ and $A_{\bf
d}({\omega_{abs}'''})B_{\bf c}({\omega_{abs}'''})$ respectively.
Note that now Mermin's claim (b) is false because the sum of the
products of ``results"
\begin{equation}
A_{\bf a}({\omega_{abs}})B_{\bf b}({\omega_{abs}}) +
A_{\bf a}({\omega_{abs}'})B_{\bf c}({\omega_{abs}'}) +
A_{\bf d}({\omega_{abs}''})B_{\bf b}({\omega_{abs}''}) -
A_{\bf d}({\omega_{abs}'''})B_{\bf c}({\omega_{abs}'''}) \label{feb25n1}
\end{equation}
does not make any mathematical sense particularly owing to theorem
1. Only if we can find a unified new abstract probability space, say
with elements $\omega_{abs}^{uni}$ do we have a justification for
the algebra of random variables and obtain:
\begin{equation}
A_{\bf a}({\omega_{abs}^{uni}})B_{\bf b}({\omega_{abs}^{uni}}) +
A_{\bf a}({\omega_{abs}^{uni}})B_{\bf c}({\omega_{abs}^{uni}}) +
A_{\bf d}({\omega_{abs}^{uni}})B_{\bf b}({\omega_{abs}^{uni}})
-A_{\bf d}({\omega_{abs}^{uni}})B_{\bf c}({\omega_{abs}^{uni}}) =
\pm 2 \label{feb25n2}
\end{equation}
as well as $M(\gamma^{j}) \leq 2$. The problem with
Eq.(\ref{adjan253}) is that Mermin's ``results" in the different
sums of the equation deserve from the start all different indices
because they may all correspond, at least in general, to different
instrument variables as well as different source variables and we
may therefore not conclude that:
\begin{equation}
M(\gamma^{j}) = {\frac {1} {4J}} \sum_{j=1}^{4J} [a_{\bf a}^j \cdot
b_{\bf b}^j + a_{\bf a}^j \cdot b_{\bf c}^j + a_{\bf d}^j \cdot
b_{\bf b}^j -  a_{\bf d}^j \cdot b_{\bf c}^j] \leq 2
\label{march06n1}
\end{equation}
The mathematically unbecoming practice of Mermin to equate or
disregard indices for objects with different origins and meaning has
led here to misconceptions and puzzlement. To put it simple: the
components of the different sums are even from a completely
classical view (without quantum non-locality) like apples and
oranges. All the parts of any given sum are just of one kind and can
be operated on with standard sampling arguments. However, one can
not mix the elements of the different sums because these are arising
from incompatible experiments. We know that we can not necessarily
add elements of different sets and therefore also not necessarily
functions on sets that differ in their very nature. Naturally, we do
not know the precise nature of EPR's elements of reality and can not
necessarily add functions of $\lambda_{\bf a}, \lambda_{\bf b}$ and
functions of the possibly completely different $\lambda_{\bf d},
\lambda_{\bf c}$. Kolmogorov's probability theory has some
flexibility: as long as at least an abstract common Kolmogorov space
can be found, we can do algebra with the functions on that space;
{\it but such common abstract space can not always be found}. Only
under certain conditions can we combine or concatenate different
Kolmogorov spaces of incompatible experiments. However, for the
particular case we consider here the enforcement of one common
domain actually changes the range (the codomain) of functions such
as $\gamma_{exp}^{j,i,l,s}$ which in turn makes a transition to a
different sample space. Naturally that leads to contradictions. The
author is very sympathetic with anybodies ``...distaste for being
explicit about what should be obvious" \cite{mermin1}. He is also
aware, however, that the great theorems of mathematics and science
have not been proven by apotheosis but by Weierstrassian generality
and rigor and sometimes tedium. The Bell theorem and the variations
by Bell's followers have not been proven with rigor and generality
but often by using simplistic examples that remind the author of the
many short ``proofs" of Fermat's theorem that have falsely claimed
to be general. The following is included to highlight this point.

We have shown in \cite{hpfound1} that the extended counterfactual
reasoning that Leggett and others have used to validate
Eq.(\ref{n08j4a}) is based on faulty logic. They justify the
appearance of different settings with labels signifying the same
entangled pair by the fact that for that given pair one could have
chosen different settings. One may indeed assume, as EPR did, that
one could have chosen a different setting in a given station to
perform the measurement, but one can not assume, as Leggett and
Peres do, but EPR did definitely not, that all these possible
results that could have been found if different settings were chosen
are then somehow found in the actual data and can be used as
elements of a sample space. One can assume that one could have
picked different items from different menus in different restaurants
but it is ridiculous to assume that one has therefore eaten them all
simultaneously. There simply exist classical experiments that are
mutually exclusive as is the appearance of certain measurement
outcomes in the actual data. Peres did acknowledge the dangers of
counterfactual arguments by stating that ``unperformed experiments
have no results". However, this statement of Peres is extremely
misleading because, taken at face value, it forbids the use of
probability theory altogether: Kolmogorov's probability theory uses
the expression $P(F)$ for the probability that an event $F$ will
occur for an experiment {\it yet to be performed} \cite{willi}.

Perhaps it is instructive to consider measurements with one given setting pair
only. Then one could have measured a quadruple with the same setting
and therefore
\begin{equation}
\gamma' =  a_{\bf a}^j \cdot b_{\bf b}^j + a_{\bf a}^j \cdot b_{\bf
b}^j + a_{\bf a}^j \cdot b_{\bf b}^j + a_{\bf a}^j \cdot b_{\bf b}^j
= \pm 4 \label{ja2801}
\end{equation}
However, such $\gamma'$ that also can be generated by the
Leggett/Peres counterfactual argument does not necessarily describe
the actual experimental outcomes in an adequate way. Consider, for
example, a classical stochastic process with time dependent $A, B$.
What does $\gamma' = \pm 4$ then mean? We could also have chosen an
n-tuple of $a_{\bf a}^j \cdot b_{\bf b}^j $ and argued that this
leads to the result $\pm n$ in Eq.(\ref{ja2801}) which depending on
the number $n$ deviates by arbitrary large amounts from the actual
sum of $n$ experimental results that may have an arbitrary time
dependence!

If, however, to save the day for Eq.(\ref{n08j4a}) Bell's $\lambda$,
and cyclically arranged functions of $\lambda$ are invoked, then we
are back full circle to the arguments of Bell and to our discussion
and refutation presented in section 3. What does then
Eq.(\ref{n08j4a}) really mean? For this author it is only proof for
how far-reaching Bell's assumptions of one common probability space
together with the cyclicity of the involved functions really are.
From the above description with indices or, a fortiori, with
measurement times $t_j,t_i,t_l,t_s$ it can be seen that
Eq.(\ref{n08j4a}) and Eq.(\ref{n08j1})  imply $J$ equalities of the
form $a_{\bf a}^j = a_{\bf a}^i$, $b_{\bf c}^i = b_{\bf c}^s$ etc.,
with $J$ being a very large integer or even considering the limit $J
\rightarrow \infty$.

That leaves only (conditional) independence arguments to possibly
justify Mermin's and Leggetts line of reasoning. These arguments are
also neither obvious nor general and are discussed here last. Bell's
demonstration and Mermin's reiteration of \cite{mermin1}
(conditional) stochastic independence of the two measurement
stations $S_1, S_2$, or the townships of Lille and Lyons in Bell's
example, suffers from the problem that it is either not general or,
if taken as general, then it is an assumption not a proof. Bell
claims that the probabilities, taken conditional to $\lambda$, that
$A_{\bf a}( (\cdot))$ in station $S_1$ and $B_{\bf b}( (\cdot))$ in
station $S_2$ assume certain values obey \cite{bellbook1}
\begin{equation}
P(A_{\bf a}( (\cdot)), B_{\bf b}( (\cdot))| \lambda) = P_1(A_{\bf a}
( (\cdot))| \lambda)P_2(B_{\bf b}( (\cdot))| \lambda)
\label{ja291}
\end{equation}
Here $(\cdot)$ indicates functional dependencies on $\lambda$ and
additional variables such as the instrument parameters described in
section 3.2. If we take the view that $\lambda$ can be anything (and
this is actually what Bell seems to do when discussing Lille and
Lyons) then the probabilities that $A, B$ take on certain values are
independent by definition. Relating to the EPR experiment that we
consider, it is inconceivable to the author that one can make all
classical physics processes in Lille and Lyons stochastically
independent by conditioning the probabilities on some ``parameter"
$\lambda$ that is being sent from some source. Why should all the
clocks of the two cities, even if set by many different and partly
confused persons and even if they are fast or slow, be
stochastically independent conditional to some entities that are
sent to the cities? Does Bell wish to tell us that all classical
physics processes in arbitrary cities can be seen as stochastically
independent conditionally to $\lambda$ and only quantum non-locality
will lead to exceptions? All of this just shows that one can not
proof theorems in such colloquial terms. One needs to agree on the
use of a respected mathematical framework. If we agree to use
Kolmogorov probability then the conditioning of the probabilities
must refer to events (subsets) of one common sample space and we
must use in our considerations one common probability space
\cite{willi}. Of course, this returns us to the (often impossible;
also classically impossible!) assumption of one common probability
space for a large number of functions that originally are defined on
a large number of different domains and correspond to incompatible
experiments.

\section{Conclusion}

The author has shown that Bell's theorem and the no-go variations of
CHSH, Leggett, Mermin, Peres and others are based on the use of one
common probability space that can not be justified for problems of
physics, classical and/or quantum when incompatible experiments are
involved. The combination of Bell inequalities and the results of
Aspect type experiments does therefore not disprove the EPR
hypothesis of the existence of elements of reality plus Einstein
locality. Furthermore, the variations on Bell by Peres, Leggett and
Mermin have been shown to neither relate to the actual experiments
nor to a valid pre-statistics (Kolmogorov's probability theory which
has a logical, clear and well defined relation to the actual
experiments) and suffer otherwise from the same problems as Bell's
theorem. The author challenges his NAS-colleagues to present a
logically rigorous and scientifically acceptable proof for the
``theorem of Bell" that encompasses all its necessary steps and uses
a valid pre-statistics with accepted relations to the actual
experiments. The author conjectures from the above discussion that
the only thing that is really impossible here is to present such a
scientific proof.

Acknowledgement: The author wishes to thank Louis Marchildon and
Peter Morgan for many valuable suggestions to improve the
manuscript.

\end{document}